\documentclass[12pt]{article}
\usepackage{graphicx}
\usepackage{amsmath}
\usepackage{amssymb}
\usepackage{latexsym}
\begin{document}
%\begin{flushright}
%BHU-PHYS-CAS Preprint\\
%arXiv: 1005.5067 [hep-th]
%\end{flushright}
\vskip 2cm
\begin{center}
{\sf {\Large{Superfield approach to nilpotent symmetries in 3D Jackiw-Pi\\ model of massive non-Abelian theory}}}

\vskip 2.0cm

{\sf S. Gupta$^{(a)}$\footnote{Present address: The Institute of Mathematical Sciences, 
CIT Campus,\\$~~~~~~~~~~~~~~~~~~~~~~~~~~~~~$Taramani, Chennai-600 113, Tamilnadu, India.}, 
R. Kumar$^{(a)}$\footnote{Present address: S. N. Bose National Centre for Basic Sciences, 
Salt Lake,\\$~~~~~~~~~~~~~~~~~~~~~~~~~~~~~$Kolkata-700098, West Bengal, India.}, R. P. Malik$^{(a,b)}$}\\
$^{(a)}$ {\it Physics Department, Centre of Advanced Studies,}\\
{\it Banaras Hindu University, Varanasi - 221 005, (U.P.), India}\\

\vskip 0.1cm

{\bf and}\\

\vskip 0.1cm

$^{(b)}$ {\it DST Centre for Interdisciplinary Mathematical Sciences,}\\
{\it Faculty of Science, Banaras Hindu University, Varanasi - 221 005, India}\\
{\small E-mails: saurabh@imsc.res.in; raviphynuc@gmail.com; rpmalik1995@gmail.com}
\end{center}

\vskip 2cm

\noindent
{\bf Abstract:} In the available literature, only the Becchi-Rouet-Stora-Tyutin (BRST) symmetries 
are known for the Jackiw-Pi model of the three (2 + 1)-dimensional (3D) massive non-Abelian gauge 
theory. We derive the off-shell nilpotent $(s_{(a)b}^2 = 0)$ and absolutely anticommuting 
$(s_b \,s_{ab} + s_{ab}\, s_b = 0)$  (anti-)BRST transformations $s_{(a)b}$ corresponding to the 
{\it usual} Yang-Mills gauge transformations of this model by exploiting the ``augmented" superfield 
formalism where the horizontality condition and gauge invariant restrictions blend together in a 
meaningful manner. There is a non-Yang-Mills (NYM) symmetry in this theory, too. However, we do not touch
the NYM symmetry in our present endeavor. This superfield formalism leads to the derivation of an 
(anti-)BRST invariant Curci-Ferrari restriction which plays a key role in the proof of absolute 
anticommutativity of $s_{(a)b}$. The derivation of the proper anti-BRST symmetry transformations is 
important from the point of view of geometrical objects called gerbes. A novel feature of our present 
investigation is the derivation of the (anti-)BRST transformations for the auxiliary field $\rho$ 
from our superfield formalism which is {\it neither} generated by the (anti-)BRST charges {\it nor} 
obtained from the requirements of nilpotency and (or) absolute anticommutativity of the (anti-)BRST 
symmetries for our present 3D non-Abelian 1-form gauge theory.\\



\newpage

\section{Introduction}
The 4D (non-)Abelian 1-form gauge theories are at the heart of standard model
of particle physics where there is a stunning degree of agreement between theory
and experiment. One of the weak links of SM is connected with the very existence 
of the Higgs particle, which is responsible for the mass generation
of gauge bosons  and fermions. In view of the fact that
Higgs particle has not yet been observed experimentally with a hundred percent certainty, other theoretical 
tools for the mass generation of gauge bosons (in various dimensions of space-time) have become 
important and they have generated a renewed interest in the realm of theoretical physics. From
many angles, the mass generation in gauge theory is an important issue.

In the above context, it may be mentioned that the 4D topologically massive
(non-) Abelian gauge theories have been studied in the past \cite{1,2,3,4} where there is
merging and mixing of 1-form and 2-form (non-)Abelian gauge fields through
the celebrated topological $(B \wedge F)$ term. In such models, it has been shown that
the (non-)Abelian 1-form gauge field acquires a mass in a very natural fashion 
without taking any recourse to the Higgs mechanism. However, these models suffer from the
problems, connected with renormalizability, consistency and unitarity. We
have studied  these models \cite{5,6,7,8,9,10}, within the framework of superfield and 
Becchi-Rouet-Stora-Tyutin (BRST) formalisms, in the hope that we would be able
to propose a model that would be free of the drawbacks of the earlier models \cite{1,2,3,4}.
However, it remains still an open problem to construct a 4D consistent, unitary and renormalizable
non-Abelian 2-form gauge theory (where the 1-form and 2-form 
non-Abelian gauge fields are incorporated together).

In this scenario, it is an interesting idea to propose and study some lower dimensional models which
are free of the problems of 4D topologically massive theory and where mass
and gauge-invariance co-exist together.
One such {\it massive} model (that has been a topic of theoretical interest) is the Jackiw-Pi 
(JP) model in three (2 + 1)-dimensions of space-time where the non-Abelian gauge 
invariance and parity are respected together due to the introduction of a 1-form vector 
field, endowed with a parity, that is opposite of the usual non-Abelian 1-form vector 
field \cite{11}. In fact, the 3D gauge theories, in general, 
have been topic of theoretical research in the recent past because of the novel and attractive properties
associated with them \cite{12,13}. Furthermore, it has already been shown 
that, the vector coupling being sufficiently strong, the gauge invariance does {\it not}
necessarily imply the {\it masslessness} of gauge particles \cite{14,15}.

Against the backdrop of these  statements, the JP model of 3D massive gauge theory has been 
studied from different theoretical angles. For instance, the Hamiltonian formulation and its 
constraint analysis have been carried out in ref. 16. The JP model is also endowed with some 
interesting continuous symmetries. In this context, mention can be made of the usual Yang-Mills 
(YM) symmetry transformation and a symmetry  that is different (i.e., non-Yang-Millg; NYM) from the YM. The 
BRST symmetry and corresponding Slavnov-Taylor identity  of this model have also been recently  
found \cite{17}. However, the off-shell nilpotent and absolutely anticommuting anti-BRST symmetry 
transformations of this model have {\it not} been discussed in ref. 17,
 which are {\it essential for the completeness} 
of the theory as their very existence is theoretically  guided and governed by 
the concept of mathematical objects called gerbes (see, e.g., refs. 18, 19 for details).

The local and continuous gauge symmetry, generated by the first-class constraints of a given
gauge theory, is generalized to the nilpotent BRST and anti-BRST symmetry transformations within
the realm of 
the BRST formalism. The anti-BRST symmetry is a new kind of symmetry transformation \cite{20} that 
is satisfied by the YM theory. It has also been shown \cite{21} that the anti-BRST symmetry 
has not been a matter of choice rather it has real fundamental importance in providing necessary 
additional conditions for the ghost freedom (that is essential for a consistent quantization). 
Both the nilpotent symmetries have been formulated in a completely  independent way  \cite{22}. 
In our recent works \cite{18,19}, we have demonstrated the relevance of gerbes in the context of BRST 
formalism through the existence of Curci-Ferrari (CF)-type restrictions. We have claimed that the 
latter is the hallmark of a gauge theory within the framework of BRST formalism. Thus, for the 
{\it sake of completeness} of the BRST analysis of the JP model, it is essential to derive a  proper 
anti-BRST symmetry corresponding to the BRST symmetry of ref. 17.

The main motivation behind our present investigation is to derive the full set of proper (i.e., off-shell
nilpotent ($s_{(a)b}^2 = 0 $) and absolutely anticommuting $(s_b s_{ab} + s_{ab} s_b = 0)$) (anti-)BRST 
symmetry transformations $s_{(a)b}$ corresponding to the usual YM gauge symmetry transformation 
for the JP model by exploiting the ``augmented" superfield approach to BRST formalism \cite{23,24,25,26,27}. 
This geometrical approach  leads to the derivation of (anti-)BRST invariant CF condition \cite{28}, 
which ensures the absolute anticommutativity of $s_{(a)b}$ and derivation of the coupled 
(but equivalent) Lagrangian densities that respect the preceding (anti-)BRST symmetry transformations
in a clear fashion.\footnote{ A more general set up for the BRST analysis of a general gauge system
exists \cite{29,30,31} within the framework of  superfield formalism where the solution of the master
equation (see, e.g., ref. 32) plays an important role. A subclass of the gauge theories, where the gauge
algebra is closed, corresponds to the Yang-Mills theories. Thus, our present endeavor could be
thought of as an application of the general approach \cite{29,30,31,32} to a specific 3D non-Abelian gauge system
with a closed gauge algebra.} Our BRST analysis, corresponding to YM symmetry transformations, 
would be complete in itself
because it is independent of the NYM symmetry transformations present in the theory.

In our present endeavor, 
we have purposely concentrated {\it only} on the usual YM gauge symmetries for the BRST analysis
within the framework of superfield formalism. This is because we plan to understand
the JP model step-by-step so that we can gain deep insights into the key aspects of this model. 
This understanding, perhaps, would enable us to propose an accurate model for the 4D theory and
would make us confident about the limiting cases of the general BRST analysis of this model where
YM and NYM gauge symmetries would be combined together for their thorough discussions. At this
juncture, it is pertinent to point out that the BRST analysis corresponding to the NYM gauge symmetry
has already been carried out in ref. 33. We have followed an
exactly similar kind of program for
the BRST analysis of the 4D dynamical non-Abelian gauge theory \cite{2,3} within the framework of
superfield approach (see, e.g., ref. 10).

The prime factors that are responsible for our present investigations are as follows. First, the JP
model is guessed to be free from the problems of unitarity and renormalizability 
encountered in the 4D topologically massive models with $(B \wedge F)$
term at the quantum level. Second, this 3D model does not invoke any higher 
form gauge field (like the 2-form $B$ field
of 4D theory) for the mass generation. Third, the understanding of this 3D theory might provide 
some insights that would show us the correct path to construct a renormalizable
and consistent 4D massive theory. Fourth, we study JP model within the framework of BRST formalism
where the renormalizability and unitarity could be proven with the help of Slavnov-Taylor
identities and {\it nilpotency} of the {\it conserved} BRST charge. 
Finally, the 3D JP model is a very special model because
it generates mass for the gauge field without violating the parity symmetry. This feature is 
drastically different from the mass generation by incorporating the Chern-Simons  term in the Lagrangian 
density of the 3D Chern-Simons  gauge theory where the parity symmetry is violated.

The contents of our present investigation are organized as follows. In Sect. 2, we discuss two sets 
of local gauge symmetry transformations associated with the JP model. Section 3 incorporates the derivation of 
(anti-)BRST symmetry transformations for the gauge field ($A_\mu$) and (anti-)ghost fields $(\bar C) C$ 
with the help of Bonora-Tonin's superfield formalism \cite{23,24}.
In Sect. 4, we deal with the derivation of (anti-)BRST symmetry 
transformations for the vector field $\phi_\mu$ and scalar field $\rho$ within the framework of ``augmented"
superfield formalism \cite{25,26,27}.
Section 5 is fully devoted to the derivation of coupled Lagrangian densities that respect the 
preceding (anti-)BRST transformations. We show the conservation of (anti-)BRST current 
(and corresponding charges) in Sect. 6. Section 7 contains the discussion of   
ghost symmetry transformations and the derivation of the algebra satisfied by {\it all} the symmetry 
generators. Finally, we made a few concluding remarks in Sect. 8.

In our Appendix A, we capture the off-shell nilpotency and  absolute anticommutativity of the
(anti-)BRST charges and the BRST invariance (as well as equivalence) of
the coupled Lagrangian densities
within the framework of ``augmented'' superfield formalism. We provide a geometrical interpretation
(through a simple diagram)
for the existence of the (anti-)BRST invariant CF restriction \cite{28} in our Appendix B with a few  
clear and cogent physical arguments.

\subsection{Conventions and notations}
 We adopt here the convention and  notations such
that the background space-time Minkowskian flat metric has the signature $(-1, +1, +1)$, totally 
antisymmetric Levi-Civita  tensor $\varepsilon_{\mu\nu\eta}$ satisfies
$\varepsilon_{\mu\nu\eta}\,\varepsilon^{\mu\nu\eta} = - 3!,\;$ $ 
\varepsilon_{\mu\nu\eta}\,\varepsilon^{\mu\nu\kappa} = - 2! \;\delta^\kappa_\eta,$ etc., 
and $\varepsilon_{012} = + 1 = - \varepsilon^{012}$. The  Greek indices 
$\mu, \nu, \eta,... = 0, 1, 2$  correspond to the 3D time and space directions. 
We take the dot and cross products $P \cdot Q = P^a \,Q^a, \;P \times Q = f^{abc}\, P^a\, Q^b\, T^c$ 
in the $SU(N)$ Lie algebraic space where the generators $T^a$ of the $SU(N)$ Lie algebra satisfy 
the commutator $[T^a, T^b] = i f^{abc}\, T^c$ with $a, b, c,... = 1, 2, 3,..., N^2 - 1$. The structure 
constants $f^{abc}$ are chosen to be totally antisymmetric in $a, b, c$ for 
the semi-simple $SU(N)$ Lie algebra \cite{34}.

\section{Preliminaries: continuous local gauge symmetries}
We begin with the Lagrangian density of the three (2 + 1)-dimensional (3D) {\it massive}
non-Abelian 1-form gauge theory proposed by JP where the $SU(N)$
YM gauge invariance and parity are respected {\it together}. Furthermore, this theory 
respects a NYM gauge symmetry transformation as well. The Lagrangian
density of the theory, in its full blaze of glory, is (see, e.g., ref. 11)
\begin{eqnarray}
{\cal L}_0 &=& - \frac{1}{4}\; F^{\mu\nu} \cdot F_{\mu\nu} - \frac{1}{4}\; \big(G^{\mu\nu} 
+ g\; F^{\mu\nu} \times \rho\big) \cdot \big(G_{\mu\nu} 
+ g \;F_{\mu\nu} \times \rho\big)\nonumber\\
&+& \frac {m}{2}\;\varepsilon^{\mu\nu\eta} \; F_{\mu\nu} \cdot \phi_\eta
\end{eqnarray}
where $A_\mu$ and $\phi_\mu$ are the vector fields with opposite parity, $\rho$ is  a scalar field, $g$ is a 
coupling constant and $m$ is the mass parameter. The 2-form ($F^{(2)} =  \frac {1}{2!} 
(dx^\mu \wedge dx^\nu)\, F_{\mu\nu} \cdot T$) curvature tensor $F_{\mu\nu} = \partial_\mu A_\nu 
- \partial_\nu A_\mu - g \;(A_\mu \times A_\nu)$, corresponding to the 1-form 
$(A^{(1)} = dx^\mu A_\mu \cdot T)$ field $A_\mu$, is derived from the 
Maurer-Cartan equation $F^{(2)} = d A^{(1)} + i\, g \;(A^{(1)} \wedge A^{(1)})$. Similarly, the field strength tensor 
$G_{\mu\nu} = D_\mu \phi_\nu - D_\nu \phi_\mu$, corresponding to the 1-form 
$(\phi^{(1)} = dx^\mu\, \phi_\mu \cdot T)$ field $\phi_\mu$, is obtained from 
$G^{(2)} = d \phi^{(1)} + i\, g \;[\phi^{(1)} \wedge A^{(1)} 
+   A^{(1)} \wedge \phi^{(1)}] \equiv \frac {1}{2!} (dx^\mu \wedge dx^\nu)\, G_{\mu\nu} \cdot T$ where the 
covariant derivative is defined as: $D_\mu \phi_\nu = \partial_\mu \phi_\nu - g \;(A_\mu \times \phi_\nu)$.

The above Lagrangian density (1) respects the following usual local, continuous and infinitesimal 
YM gauge transformations $\delta_1$, as 
\begin{eqnarray}
&&\delta_1 A_\mu = D_\mu \Lambda \qquad\qquad \delta_1 \phi_\mu = - g\,(\phi_\mu \times \Lambda) \qquad\qquad
\delta_1 \rho = -\, g\,(\rho \times \Lambda) \nonumber\\
&&\delta_1 F_{\mu\nu} = - \,g\,(F_{\mu\nu} \times \Lambda) \qquad 
\qquad \delta_1 G_{\mu\nu} = - \,g\,(G_{\mu\nu} \times \Lambda)
\end{eqnarray}
This theory also respects a NYM gauge transformation $\delta_2$.  The
infinitesimal version of this transformation is
\begin{eqnarray}
&&\delta_2 A_\mu = 0 \qquad  \qquad \delta_2 \phi_\mu = D_\mu \Omega \qquad  \qquad \delta_2 \rho 
= +\; \Omega  \nonumber\\
&& \delta_2 F_{\mu\nu} = 0 \quad \qquad \delta_2 G_{\mu\nu} 
= - \,g\,(F_{\mu\nu} \times \Omega)
\end{eqnarray} 
where $\Lambda = \Lambda \cdot T \equiv \Lambda^a \;T^a$ and $\Omega = \Omega \cdot T \equiv 
\Omega^a\; T^a$ are the $SU(N)$ valued infinitesimal (Lorentz-scalar)  gauge parameters. 
It is straightforward to check that the Lagrangian density (1) transforms, under the local, continuous 
and infinitesimal transformations (2) and (3), as 
\begin{eqnarray}
\delta_1 {\cal L}_0 = 0 \quad \qquad\qquad \delta_2 {\cal L}_0 = \partial_\mu \Big[\frac {m}{2}\; 
\varepsilon^{\mu\nu\eta} \;F_{\nu\eta} \cdot \Omega \Big]
\end{eqnarray}
We note that the usual YM symmetry is a perfect symmetry for ${\cal L}_0$ because we have perfect invariance
(i.e., $\delta_1 {\cal L}_0 = 0$). However, the Lagrangian density ${\cal L}_0$ remains quasi-invariant 
under $\delta_2$ because it transforms to a total space-time derivative. Furthermore, we observe
 that the YM and
NYM symmetries (i.e., $\delta_1$ and $\delta_2$) are independent of each-other \cite{11,12,13}.

In our present investigation, we shall concentrate only  on the {\it usual} YM gauge symmetry transformations 
$(\delta_1)$ and derive  the corresponding (anti-)BRST symmetry transformations that would be off-shell 
nilpotent and absolutely anticommuting in nature. In other words, we shall demonstrate that our BRST analysis 
of YM symmetry transformations would be 
{\it complete} in itself because, as we have stated earlier,
the transformations   $\delta_1$ and $\delta_2$
are independent of each other. The BRST analysis, corresponding to the NYM gauge transformations has been
performed by two of us \cite{33}, about which we mention concisely in Sect. 8.

\section{ (Anti-)BRST symmetries of gauge and (anti-)ghost fields: Bonora-Tonin's superfield formalism}

We apply the well-known Bonora-Tonin's superfield approach \cite{23,24}
to derive the nilpotent (anti-)BRST symmetry 
transformations corresponding to the YM gauge transformations ($\delta_1$) for the 1-form 
gauge field $A_\mu$ and (anti-)ghost fields $(\bar C)C$. In this approach, first of all, we generalize the 3D 
bosonic vector field $(A_\mu = A_\mu \cdot T)$ and fermionic (anti-)ghost fields $(\bar C = \bar C 
\cdot T, \; C =  C \cdot T)$ to their corresponding superfields. The latter are defined on the 
$(3, 2)$-dimensional supermanifold parametrized by the superspace variables $Z^M = (x^\mu, \theta, \bar \theta)$
where $x^\mu (\mu = 0, 1, 2)$ are the space-time variables and $(\theta, \bar \theta)$ are a pair of Grassmannian
variables (with $\theta^2 = \bar \theta^2 = 0, \; \theta \,\bar \theta + \bar \theta\, \theta = 0$). These superfields 
can be expanded along the Grassmannian 
directions  $\theta$ and $\bar \theta$ (of the (3, 2)-dimensional supermanifold), as \cite{23,24}
\begin{eqnarray}
\tilde B_{\mu} (x, \theta, \bar\theta) &=& A_\mu (x) + \theta\; \bar R_\mu (x) + \bar \theta\; R_\mu (x)
+ i \;\theta \;\bar\theta \; S_\mu (x) \nonumber\\
\tilde F (x, \theta, \bar\theta) &=& C (x) + i\;\theta\; \bar B_1 (x) + i\;\bar \theta\; B_1 (x)
+ i \;\theta\; \bar\theta \; s (x) \nonumber\\
\tilde {\bar F} (x, \theta, \bar\theta) &=& \bar C (x) +  i\;\theta\; \bar B_2 (x) + i\;\bar \theta\; 
B_2 (x) + i \;\theta \;\bar\theta \; \bar s (x)
\end{eqnarray}
where the local secondary fields $[R_\mu (x), \, \bar R_\mu (x), \, s(x),$ $\, \bar s(x)]$ are fermionic 
($s^2 = 0,\, \bar s^2 = 0, \,R_\mu R^\mu = 0,$ $ \, R_\mu \bar R_\nu + \bar R_\nu R_\mu = 0,$ etc.) 
and $[S_\mu (x),\;  B_1 (x),\; \bar B_1 (x), \; B_2 (x),\;$  $\bar B_2 (x)]$ are 
bosonic in nature. These secondary fields can be determined in terms of the basic and auxiliary 
fields of the 3D local BRST invariant quantum field theory through the application of the celebrated 
horizontality condition (HC).

We  note that the kinetic term $[- \frac {1}{4}F^{\mu\nu} \cdot F_{\mu\nu}]$, corresponding 
to the 1-form gauge field $A_\mu$, remains invariant under the gauge transformations (2). The HC 
condition implies that the gauge invariant kinetic term remains invariant when we generalize the
3D ordinary non-Abelian gauge theory onto (3, 2)-dimensional supermanifold. The above statement 
of the gauge invariance can be, mathematically, expressed as: 
\begin{eqnarray}
 - \frac {1}{4} \; F^{\mu\nu} \cdot F_{\mu\nu} = - \frac {1}{4} {\tilde F}^{MN} \cdot {\tilde F}_{MN}
\end{eqnarray}
where the super curvature ${\tilde F}^{MN}$, defined on the $(3, 2)$-dimensional supermanifold, 
is derived from the Maurer-Cartan equation ${\tilde F}^{(2)} = \tilde d {\tilde A}^{(1)} 
+ \,i\, g\;({\tilde A}^{(1)} \wedge {\tilde A}^{(1)}) 
\equiv  \frac {1}{2!} \,(dZ^M \wedge dZ^N)\,{\tilde F}_{MN}$. Here $\tilde d$ is the super exterior derivative
(with ${\tilde d}^2 = 0$) and $\tilde {A^{(1)}}$ is the super 1-form connection which 
are the generalizations of the ordinary exterior 
derivative $d$ and 1-form connection $A^{(1)}$ as 
\begin{eqnarray}
 d \longrightarrow  \tilde d &=& dZ^M \; \partial_M 
 \equiv dx^\mu \;\partial_\mu + d \theta \;\partial_\theta
+ d \bar \theta \;\partial_{\bar\theta} \nonumber\\
 A^{(1)} \longrightarrow   \tilde A^{(1)} &=& dZ^M \tilde A_M \nonumber\\
&\equiv& dx^\mu \;\tilde B_\mu (x,\theta,\bar\theta) + d \theta\;
\tilde {\bar F} (x,\theta,\bar\theta) + d \bar \theta \;\tilde F (x,\theta,\bar\theta)
\end{eqnarray}
where $\tilde B_\mu (x, \theta, \bar \theta),\, \tilde F(x, \theta, \bar \theta)$ and 
$\tilde {\bar F}(x, \theta, \bar \theta)$ are the superfields defined on the (3, 2)-dimensional 
supermanifold and $\partial_M = (\partial_\mu, \partial_\theta, \partial_{\bar \theta})$. 
The celebrated HC condition (6) leads to the following 
relationships amongst the basic, auxiliary and secondary fields \cite{23,24}
\begin{eqnarray}
R_\mu &=& D_\mu C \qquad \bar R_\mu = D_\mu \bar C \qquad B_1 = -  \frac{i}{2}\;g\;(C \times C)\qquad
\bar B_2 = - \frac{i}{2}\; g\;(\bar C \times \bar C) \nonumber\\
s &=& -\,g\;(\bar B \times C) \qquad \bar s = +\, g\;(B \times \bar C) \qquad  B + \bar B = -\,i\, g\;(C \times \bar C)\nonumber\\
S_\mu &=& D_\mu B \;+ \;i\, g \;(D_\mu C \times \bar C) \equiv  - \,D_\mu \bar B - \;i\, g\;(D_\mu \bar C \times C) 
\end{eqnarray}
where we have made the choices $\bar B_1 = \bar B$ and $B_2 = B$ which are, finally, 
identified with the Nakanishi-Lautrup type auxiliary fields of the 3D local quantum field theory
within the framework of BRST formalism.

It is to be noted that, to satisfy the HC (6), one sets equal to zero the Grassmannian components 
of the super tensor $\tilde F_{MN}$ in super 2-form $\tilde F^{(2)} = \frac{1}{2}\big(dZ^M \wedge d Z^N\big) 
\tilde F_{MN}$. The equation  $\bar B + B = - \,i\, g \; (C \times \bar C)$ (quoted in (8)) is the 
CF restriction, which is one of the key {\it hallmarks} of the non-Abelian 1-form gauge theory. 
This condition is derived from HC when one sets equal to zero the $\tilde F_{\theta \bar\theta}$ component 
of the super curvature tensor $\tilde F_{MN}$. The CF 
condition plays an important role in providing the proof for the absolute anticommutativity of the (anti-)BRST 
transformations. Furthermore, the CF condition is instrumental in obtaining a coupled set of Lagrangian
densities (see (30) and (31) below) that respect the (anti-)BRST transformations (see also Appendix B).

Substituting the above relationships (8) into the super-expansions of the superfields, (5), 
we obtain the following explicit expansions:  
\begin{eqnarray}
\tilde B^{(h)}_{\mu} (x, \theta, \bar\theta) &=& A_\mu (x) + \theta\, D_\mu \bar C (x) 
+ \bar \theta\, D_\mu C (x) +  \theta \,\bar\theta \, [i\, D_\mu  B - g\,(D_\mu C \times \bar C)] (x) \nonumber\\
&\equiv& A_\mu (x) + \theta\, (s_{ab} A_\mu (x)) + \bar \theta\, (s_b A_\mu (x)) 
+ \theta\, \bar\theta\; (s_b s_{ab} \;A_\mu (x)) \nonumber\\
\tilde F^{(h)} (x, \theta, \bar\theta) &=& C (x) + \theta\; (i\, \bar B (x)) + \bar \theta\; \Bigl 
[\frac{g}{2}\; (C \times C) (x) \Bigr] + \theta \bar\theta \; [-\,i\, g\;(\bar B \times C)(x)] \nonumber\\
&\equiv& C (x) + \theta\; (s_{ab} C (x)) + \bar \theta\; (s_b C (x)) + \theta\; \bar\theta\; (s_b s_{ab} \;C (x))\nonumber\\
\tilde {\bar F}^{(h)} (x, \theta, \bar\theta) &=& \bar C (x) +
\theta\; \Bigl [\frac{g}{2} \;(\bar C \times \bar C) (x) \Bigr ] + \bar \theta\; (i  B (x))
+ \theta \bar\theta \; [(+ \,i\,g\;(B \times \bar C) (x)] \nonumber\\
&\equiv& \bar C (x) + \theta\; (s_{ab} \bar C (x)) + \bar \theta\; (s_b \bar C (x))
+ \theta\; \bar\theta\; (s_b s_{ab}\; \bar C (x))
\end{eqnarray}
where $(h)$, as the superscript on the superfields, denotes the expansions of the superfields 
after the application of HC. The super 2-form curvature tensor can be expressed  as 
\begin{eqnarray}
\tilde F^{(h)}_{\mu\nu} (x,\theta,\bar\theta) &=& F_{\mu\nu}(x) - \theta \; 
[g\; (F_{\mu\nu} \times \bar C)](x)- \bar\theta \;[g \;(F_{\mu\nu} \times C)](x) \quad\nonumber\\
&+& \theta\; \bar\theta\, \bigl [g^2\,(F_{\mu\nu} \times C) \times \bar C -\,i\,g \,(F_{\mu\nu} \times B) \bigr ](x)\quad\nonumber\\
&\equiv&   F_{\mu\nu}(x) + \theta \,(s_{ab} F_{\mu\nu}(x)) 
+ \bar\theta \,(s_b F_{\mu\nu}(x)) + \theta\, \bar\theta\, (s_b\,s_{ab}\,F_{\mu\nu}(x))
\end{eqnarray}
It is clear, from the above expressions, that the kinetic term of $A_\mu$ (i.e., $- \frac{1}{4}\; F^{\mu\nu} 
\cdot F_{\mu\nu}$) remains invariant (i.e., independent of the Grassmann variables $\theta, \bar 
\theta$) under the application of HC. In other words, we obtain 
$- (1/4) \tilde F^{\mu\nu(h)} (x, \theta, \bar\theta) \cdot \tilde F^{(h)}_{\mu\nu} (x, \theta, \bar\theta)
= - (1/4) F^{\mu\nu} (x) \cdot F_{\mu\nu} (x)$. 
Before we wrap up this section, we would like to state that the equations (9) and (10) imply that:
$s_b \Psi(x) = (\partial/ \partial \bar \theta)\, \tilde \Psi^{(h)}(x, \theta, \bar \theta) 
\big |_{\theta = 0},\, 
s_{ab} \Psi(x) = (\partial/ \partial \bar \theta)\, \tilde \Psi^{(h)}(x, \theta, \bar \theta) 
\big |_{\bar \theta = 0}$ where $\Psi(x)$ is the generic 3D ordinary field and $\tilde \Psi^{(h)}(x, \theta, \bar \theta)$ is
the corresponding superfield obtained after the application of HC. This
mapping establishes a relationship between the (anti-)BRST  transformations and the translational 
generators along the Grassmannian directions of the $(3, 2)$-dimensional supermanifold.
This key relationship entails upon the (anti-)BRST transformations, emerging 
from our superfield formalism, to be {\it always} nilpotent of order two ($s_{(a)b}^2 = 0$) and
absolutely anticommuting in nature because we have: $(\partial_\theta)^2 = (\partial_{\bar \theta})^2 
= 0$ and $\partial_\theta \;\partial_{\bar \theta} + \partial_{\bar \theta}\; \partial_\theta = 0$.

\section{(Anti-)BRST symmetries for the vector field ($\phi_\mu$) and scalar field ($\rho$) : 
augmented superfield formalism}
It can be checked that the composite fields $(F_{\mu\nu}\cdot \phi_\eta)$ and 
$ (F_{\mu\nu} \cdot \rho)$ remain 
invariant under the usual YM gauge transformations $(\delta_1)$ 
\begin{eqnarray}
\delta_1 (F_{\mu\nu}\cdot \phi_\eta) = 0 \qquad \qquad  \delta_1 (F_{\mu\nu} \cdot \rho) = 0
\end{eqnarray}
because $(F_{\mu\nu}\cdot \phi_\eta)$ and $ (F_{\mu\nu} \cdot \rho)$ are the gauge-invariant quantities. 
Therefore, these are physical quantities (in some sense). These quantities must remain unaffected by 
the presence of the Grassmannian variables when the former entities are generalized onto the (3, 2)-dimensional 
supermanifold. Thus, we have the following gauge invariant restrictions (GIRs) on the (super)fields:
\begin{eqnarray}
\tilde F_{\mu\nu}^{(h)} (x,\theta,\bar\theta) \cdot \tilde \phi_\eta (x,\theta,\bar\theta)
&=& F_{\mu\nu} (x) \cdot \phi_{\eta} (x) \nonumber\\
\tilde F_{\mu\nu}^{(h)} (x,\theta,\bar\theta) \cdot \tilde \rho (x,\theta,\bar\theta)
&=& F_{\mu\nu} (x) \cdot \rho (x)
\end{eqnarray}
where the expansion for $\tilde F^{(h)}_{\mu\nu} (x, \theta, \bar\theta)$ is quoted in equation (10).
The bosonic superfields $\tilde \phi_\mu (x,\theta,\bar\theta)$ and $\tilde \rho (x,\theta,\bar\theta)$
can be, in general, expanded on the (3, 2)-dimensional supermanifold along the Grassmannian directions ($\theta, 
\bar \theta$), as  
\begin{eqnarray}
\tilde \phi_\mu (x,\theta,\bar\theta) &=& \phi_\mu (x) + \theta\; \bar P_\mu (x) + \bar\theta\;
P_\mu (x) + i \;\theta \;\bar\theta\; Q_\mu (x)\nonumber\\
\tilde \rho (x,\theta,\bar\theta) &=& \rho (x) + \theta \; \bar P (x)
+ \bar\theta\; P (x) + i \;\theta \;\bar\theta \;Q(x) 
\end{eqnarray}  
where the secondary fields $(P_\mu, \bar P_\mu, P, \bar P)$ are fermionic and $(Q_\mu, Q)$ are bosonic 
in nature. These secondary fields can be determined with the help of GIRs, (12). 
In fact, the equality (12), leads to the following relationships:   
\begin{eqnarray}
&&P_\mu = - \,g\,(\phi_\mu \times C) \qquad\qquad \bar P_\mu = - \,g\,(\phi_\mu \times \bar C) 
\qquad\qquad P = -\, g\,(\rho \times C) \nonumber\\
&& Q_\mu = - \,i\, \bigl [g^2\,(\phi_\mu \times C) \times \bar C 
- \,i\, g\,(\phi_\mu \times B) \bigr ]   \qquad \qquad  \bar P = -\, g\,(\rho \times \bar C)\nonumber\\
&& Q = - \,i\, \bigl [g^2\,(\rho \times C) \times \bar C -\, i\, g\,(\rho \times B) \bigr ]\nonumber\\
\end{eqnarray}
Substituting  the above values into  (13), we obtain 
\begin{eqnarray} 
\tilde \phi^{(h,g)}_\mu (x,\theta,\bar\theta) &=& \phi_\mu (x) - \theta\; [g\;(\phi_\mu \times \bar C)] (x) 
- \bar\theta\; [g\;(\phi_\mu \times C)] (x) \nonumber\\ 
& +& \theta \,\bar\theta\, [g^2\,(\phi_\mu \times C) \times \bar C -\, i\, g\, (\phi_\mu \times B) ](x)\nonumber\\
&\equiv& \phi_\mu (x) + \theta \;(s_{ab}\; \phi_\mu (x)) + \bar \theta \;(s_b \;\phi_\mu (x))
+ \theta \;\bar \theta \;(s_b \;s_{ab} \;\phi_\mu (x))\nonumber\\
\tilde \rho^{(h,g)} (x,\theta,\bar\theta) &=& \rho (x) - \theta \; [g\;(\rho \times \bar C)] (x)
- \bar\theta\; [g\;(\rho \times C)] (x) \nonumber\\
&+& \theta \;\bar\theta \;[g^2\;(\rho \times C) \times \bar C - \,i\, g \;(\rho \times B)] (x)\nonumber\\
&\equiv& \rho (x) + \theta \;(s_{ab}\; \rho (x)) + \bar \theta \;(s_b \;\rho (x)) 
+ \theta \;\bar \theta \;(s_b \;s_{ab} \;\rho (x))
\end{eqnarray}
where the superscripts $(h,g)$ on the above superfields refers to the super-expansions of the superfields
obtained after the application of celebrated 
HC condition and GIRs. Thus, we have already obtained
the (anti-)BRST symmetry transformations here
for the fields $\phi_\mu$ and $\rho$ in view of the mappings:
$s_b = (\partial/ \partial \bar \theta) \big |_{\theta = 0}, \; 
s_{ab} = (\partial/ \partial \theta) \big |_{\bar \theta = 0}$. In other words, the coefficients of 
$\theta$ and  $\bar \theta,$ in these expansions, provide the (anti-)BRST symmetry transformations, respectively.

Within the framework of superfield formalism, we can also calculate the nilpotent (anti-)BRST transformations 
for the field strength tensor $G_{\mu\nu}$ and composite field ($F_{\mu\nu} \times \rho$). With the
inputs from Sect. 3 and Sect. 4, we have the following:
\begin{eqnarray}
\tilde G^{(h, g)}_{\mu\nu}& =& \partial_\mu \tilde\phi^{(h,g)}_\nu - g\; \Big(\tilde B^{(h)}_\mu \times 
\tilde\phi^{(h,g)}_\nu \Big) - \partial_\nu \tilde\phi^{(h,g)}_\mu 
+ g\; \Big(\tilde B^{(h)}_\nu \times \tilde\phi^{(h,g)}_\mu \Big)
\end{eqnarray}
The substitution of expansions from (9) and (15) leads to the following expansion:
\begin{eqnarray}
\tilde G^{(h, g)}_{\mu\nu} (x, \theta, \bar \theta)  &=&  G_{\mu\nu} (x) 
+ \theta \, [ -g \, (G_{\mu\nu} \times \bar C)]+ \bar \theta \, [ -g \; (G_{\mu\nu} \times C)](x) \nonumber\\
&+& \theta \bar \theta \, [ g^2 \, (G_{\mu\nu} \times C) \times \bar C - i\, g \, (G_{\mu\nu} \times B)](x)
\end{eqnarray}
which results in the following off-shell nilpotent 
(anti-)BRST transformations for the tensor field $G_{\mu\nu}$, namely; 
\begin{eqnarray}
&& s_b G_{\mu\nu} =   -\, g \, (G_{\mu\nu} \times C) \qquad 
\qquad s_{ab} G_{\mu\nu} = -\, g \, (G_{\mu\nu} \times \bar C)\nonumber \\
&& s_b \,s_{ab} G_{\mu\nu} =  g^2 \, (G_{\mu\nu} \times C) \times \bar C - i \,g \, (G_{\mu\nu} \times B)
\end{eqnarray}
It is also interesting to check explicitly that
\begin{eqnarray}
\tilde G^{(h, g)}_{\mu\nu} (x, \theta, \bar \theta) \cdot \tilde G^{\mu\nu (h, g)} (x, \theta, \bar \theta) 
 =  G_{\mu\nu} (x) \cdot G^{\mu\nu} (x)
\end{eqnarray}
which establishes the Grassmannian independence of the left-hand side As a consequence, we infer from this 
observation that $(G_{\mu\nu}  \cdot G^{\mu\nu})$ is an (anti-)BRST 
invariant quantity.

In an exactly similar fashion, it is straightforward to note that (cf. (10) and (15))
\begin{eqnarray}
(\tilde F_{\mu\nu}^{(h)} \times \tilde \rho^{(h,g)}) (x, \theta, \bar \theta) & =&
(F_{\mu\nu} \times \rho ) (x) - \theta \, [g\, (F_{\mu\nu} \times  \rho ) \times \bar C] (x)
- \bar \theta \, [g \,(F_{\mu\nu} \times  \rho ) \times  C] (x) \nonumber\\
&+&\theta\, \bar \theta \, [ g^2 \, \{(F_{\mu\nu} \times \rho) \times C \} \times \bar C
- i g \,  (F_{\mu\nu} \times \rho) \times B] (x)
\end{eqnarray}
This expansion  implies that the (anti-)BRST symmetry transformations of the composite 
field $(F_{\mu\nu} \times  \rho )$ are 
\begin{eqnarray}
&& s_b (F_{\mu\nu} \times  \rho ) = -\, g \, (F_{\mu\nu} \times  \rho ) \times C 
\qquad \qquad s_{ab} (F_{\mu\nu} \times  \rho ) = - \,g \, (F_{\mu\nu} \times  \rho ) \times \bar C \nonumber\\
&& s_b \,s_{ab} \; (F_{\mu\nu} \times  \rho ) = g^2 \; \{(F_{\mu\nu} \times \rho) \times C \} \times \bar C
-\, i\, g \;  (F_{\mu\nu} \times \rho) \times B
\end{eqnarray}
Furthermore, it is elementary to show that the following is correct, namely; 
\begin{eqnarray}
\Big[\Big(\tilde F_{\mu\nu}^{(h)} \times \tilde \rho^{(h,g)}\Big) (x, \theta, \bar \theta) \Big] \cdot
\Big [ \Big(\tilde F^{\mu\nu (h)} \times \tilde \rho^{(h,g)}\Big) (x, \theta, \bar \theta) \Big]
 = (F_{\mu\nu} \times  \rho ) (x) \cdot (F^{\mu\nu} \times  \rho ) (x)
\end{eqnarray}
which establishes the Grassmannian independence of the left-hand side. As a consequence, we conclude 
that $(F_{\mu\nu} \times  \rho ) \cdot (F^{\mu\nu} \times  \rho )$ is an (anti-)BRST invariant
quantity (i.e., $s_{(a)b} \; [(F_{\mu\nu} \times  \rho )  \cdot (F^{\mu\nu} \times  \rho )] = 0 $).
Before we close this section, it is interesting to note that the following equation is correct:
\begin{eqnarray}
\tilde G^{(h, g)}_{\mu\nu} (x, \theta, \bar \theta) \cdot \Big(\tilde F^{\mu\nu (h)} 
\times \tilde \rho^{(h,g)}\Big) (x, \theta, \bar \theta) \; = \; G_{\mu\nu} (x) \cdot 
(F^{\mu\nu} \times \rho) (x) 
\end{eqnarray}
It verifies the Grassmannian independence of the left-hand side. This observation, in turn, 
implies the (anti-)BRST invariance of $G_{\mu\nu}  \cdot (F^{\mu\nu} \times \rho) $
(i.e., $ s_{(a)b} \; [G_{\mu\nu} \cdot (F^{\mu\nu} \times \rho)] = 0$). Finally, it is 
clear that $\big( G^{\mu\nu} + g (F^{\mu\nu} \times \rho) \big) \cdot \big( G_{\mu\nu} 
+ g (F_{\mu\nu} \times \rho) \big) $ of the Lagrangian density (1) is
an (anti-)BRST invariant quantity. Furthermore, it is true to state that the following is correct:
\begin{eqnarray}
\frac {m}{2} \; \varepsilon^{\mu\nu\eta} \; \tilde F_{\mu\nu}^{(h)} (x, \theta, \bar\theta) \cdot 
\tilde \phi_\eta^{(h,g)} 
(x, \theta, \bar\theta) \; = \; \frac {m}{2}\;  \varepsilon^{\mu\nu\eta} \; F_{\mu\nu} (x)  \cdot \phi_\eta (x)
\end{eqnarray}
which proves the (anti-)BRST invariance of the last term of ${\cal L}_0$ because the left-hand side  of (24) is 
actually independent of the Grassmannian variables.

\section{Coupled Lagrangian densities: (anti-)BRST symmetries}

In ourprevious Sect. 4,  we have derived the (anti-)BRST symmetry transformations for 
the relevant fields of the theory. This can be seen from a close look at equations (9), (10), and (15).
In explicit form, these (anti-)BRST transformations are 
\begin{eqnarray}
&&s_{ab} A_\mu = D_\mu \bar C \qquad s_{ab} \bar C = \frac {g}{2} \,(\bar C \times \bar C) 
 \qquad s_{ab} B = - g\,(B \times \bar C) \qquad s_{ab} \rho = - \,g\,(\rho \times \bar C) \quad\nonumber\\ 
&&  s_{ab} \phi_\mu = - \,g\,(\phi_\mu \times \bar C)\qquad 
s_{ab} C = i \,\bar B \qquad s_{ab} \bar B = 0 \qquad s_{ab} F_{\mu\nu} = - \,g\,(F_{\mu\nu} \times \bar C) \qquad
\end{eqnarray}
\begin{eqnarray}
&& s_b A_\mu = D_\mu C \qquad s_b C =  \frac{g}{2}\; (C \times C) 
\qquad s_b \bar B = -\, g\,(\bar B \times C) \qquad s_b \rho = - \,g\, (\rho \times C) \qquad\nonumber\\
&& s_b \phi_\mu = - \,g\,(\phi_\mu \times C) \qquad s_b \bar C = i\, B \qquad 
s_b B = 0 \qquad  s_b F_{\mu\nu} = -\, g\,(F_{\mu\nu} \times C) \qquad
\end{eqnarray}  
These transformations are off-shell nilpotent of order two $(s_{a(b)}^2 = 0)$ and 
absolutely anticommuting $(s_b\, s_{ab} + s_{ab}\, s_b = 0)$ in nature in their operator form.

The (anti-)BRST symmetry transformations for the Nakanishi-Lautrup type auxiliary fields ($B,\, \bar B$)
have been obtained from the requirements of the nilpotency and absolute anticommutativity properties of 
$s_{(a)b}$. In fact, the above requirements lead to the following  proper (anti-)BRST  symmetry transformations: 
\begin{eqnarray}
&& s_b B = 0 \qquad  s_b \bar B = -\, g\,(\bar B \times C) \qquad  s_{ab} \bar B = 0
\qquad  s_{ab} B = - \,g\,(B \times \bar C)
\end{eqnarray}
The absolute anticommutativity of the (anti-)BRST symmetry 
transformations ($s_{(a)b}$), applied onto the following fields:
\begin{eqnarray}
&& \{s_b,\; s_{ab}\} \;A_\mu = 0 \qquad  \qquad \{s_b,\; s_{ab}\} \;\phi_\mu = 0
\qquad\qquad \{s_b,\; s_{ab}\} \;\rho = 0
\end{eqnarray}
is {\it true} only when CF-condition, $B + \bar B + i\, g\, (C \times \bar C) = 0$, is satisfied.
Thus, we note that our superfield formalism leads to (i) the derivation of the off-shell nilpotent 
and absolutely anticommuting (anti-)BRST transformations, and (ii) the CF condition. The latter 
defines a hypersurface in the 3D  Minkowski space-time manifold on which the proper (i.e., nilpotent and 
absolutely anticommuting) (anti-)BRST transformations are defined. Furthermore, it can be
checked that the well-known CF-condition is (anti-)BRST invariant (i.e., $s_{(a)b} \, [B + \bar B 
+ i\, g\, (C \times \bar C)] = 0$). In other words, the key results of our superfield formalism are the
derivation of the {\it proper} (anti-)BRST transformations and (anti-)BRST invariant CF-condition
for the BRST invariant theory. It would be worthwhile to mention here that
the (anti-)BRST symmetry transformations (25)-(27) are {\it also} consistent with the general setup
of superfield formalism developed in  ref. 30. It is, primarily, because of  the closed-algebraic   
structure of our present theory that we do not obtain the higher-order (anti-)ghost fields in the 
off-shell nilpotent symmetry transformations 
(25)-(27) (for the JP model).

The expressions for the coupled (anti-)BRST-invariant Lagrangian densities of the 
3D massive non-Abelian gauge theory can be written as 
\begin{eqnarray}
{\cal L}_B = {\cal L}_0 + s_b \, s_{ab} \bigg[\frac {i}{2} \, A_\mu \cdot A^\mu + C \cdot \bar C 
+ \frac {1}{2}\, \phi_\mu \cdot \phi^\mu \bigg]\nonumber\\
{\cal L}_{\bar B} = {\cal L}_0 - s_{ab} \, s_b \bigg[\frac {i}{2} \, A_\mu \cdot A^\mu 
+ C \cdot \bar C + \frac {1}{2}\, \phi_\mu \cdot \phi^\mu \bigg]
\end{eqnarray}
We  note that the terms in the square brackets are chosen in such a way that each term has 
a mass dimension one and ghost number equal to zero. Furthermore, all these terms (in the bracket) are 
Lorentz scalar. This is because  the proper nilpotent (anti-)BRST
symmetries increase the mass dimension of fields by {\it one} on which they operate.  
As a consequence, we have the following expressions for the 3D (anti-)BRST invariant 
coupled (but equivalent) Lagrangian densities   
\begin{eqnarray}
{\cal L}_B &=& - \frac{1}{4}\; F^{\mu\nu} \cdot F_{\mu\nu} - \frac{1}{4}\; \big(G^{\mu\nu} 
+ g\, F^{\mu\nu} \times \rho\big) \cdot \big(G_{\mu\nu} 
+ g\, F_{\mu\nu} \times \rho\big)
+ \frac {m}{2}\;\varepsilon^{\mu\nu\eta} \; F_{\mu\nu} \cdot \phi_\eta \nonumber\\
&+&  B \cdot (\partial^\mu A_\mu) + \frac {1}{2} (B \cdot B + \bar B \cdot \bar B)
- i\; \partial^\mu \bar C \cdot D_\mu C
\end{eqnarray}
\begin{eqnarray}
{\cal L}_{\bar B} &=& -  \frac{1}{4}\; F^{\mu\nu} \cdot F_{\mu\nu} - \frac{1}{4}\; \big(G^{\mu\nu} 
+ g\, F^{\mu\nu} \times \rho\big) \cdot \big(G_{\mu\nu} 
+ g\, F_{\mu\nu} \times \rho\big) 
+ \frac {m}{2}\;\varepsilon^{\mu\nu\eta} \; F_{\mu\nu} \cdot \phi_\eta \nonumber\\
&-& \bar B \cdot (\partial^\mu A_\mu) + \frac {1}{2} (B \cdot B + \bar B \cdot \bar B) 
- i\; D^\mu \bar C \cdot \partial_\mu C
\end{eqnarray}
We emphasize that there are no gauge-fixing and Faddeev-Popov ghost terms for the vector field $\phi_\mu$ 
in the Lagrangian densities (30) and (31). The reason behind this observation, is that the field 
$\phi_\mu$ transforms covariantly (i.e., $s_b \phi_\mu = -\, g\;(\phi_\mu \times C), \;
s_{ab} \phi_\mu = -\, g\;(\phi_\mu \times \bar C)$) under the (anti-)BRST transformations, (25) and 
(26). Thus, the term $(\phi_\mu \cdot \phi^\mu)$ remains invariant under the off-shell nilpotent 
(anti-)BRST transformations
and, as a consequence, there is no contribution from this term to the (anti-)BRST invariant 
coupled (but equivalent) Lagrangian densities quoted in (30) and (31).

It can be checked that the preceding  Lagrangian densities, ${\cal L}_B$ and ${\cal L}_{\bar B}$, transform, 
under the off-shell nilpotent (anti-)BRST transformations, as
\begin{eqnarray}
s_b {\cal L}_B &=& \partial_\mu[B\cdot (D^\mu C)] \qquad \qquad
s_{ab} {\cal L}_{\bar B} = - \,\partial_\mu[\bar B\cdot (D^\mu \bar C)]\nonumber\\
s_{ab} {\cal L}_B &=& \partial_\mu(B \cdot \partial^\mu \bar C) 
- D_\mu [B + \bar B + i\, g\, (C \times \bar C)] \cdot \partial^\mu \bar C \nonumber\\
s_b {\cal L}_{\bar B} &= &-\, \partial_\mu(\bar B \cdot \partial^\mu C) 
+ D_\mu [B + \bar B + i\, g (C \times \bar C)] \cdot \partial^\mu C
\end{eqnarray}
Thus, the corresponding actions (i.e., $\int d^2 x \;{\cal L}_B$ 
and $\int d^2 x \;{\cal L}_{\bar B}$) remain invariant on the constrained hypersurface in the 3D space-time
manifold where the CF-condition is satisfied. The explicit expressions in (32) explain that 
${\cal L}_B$ and ${\cal L}_{\bar B}$ are {\it equivalent} and both of them respect the nilpotent (anti-)BRST symmetry
transformations.

\section{ Conserved charges: novel features}

The invariance of the Lagrangian density (or action), under any arbitrary continuous symmetry 
transformation, leads to the derivation of the conserved current according to Noether's theorem. 
As a consequence, the local, continuous, and 
off-shell nilpotent (anti-)BRST symmetry transformations ($s_{(a)b}$) lead to 
the derivation of the following Noether's conserved currents, $J^\mu_{(a)b}$: 
\begin{eqnarray} 
J^\mu_b &=& B \cdot (D^\mu C) - \big[F^{\mu\nu} - g\,((G^{\mu\nu} 
+ g \,F^{\mu\nu} \times \rho) \times \rho) 
- m \,\varepsilon^{\mu\nu\eta} \, \phi_\eta \big]  \cdot (D_\nu C)\nonumber\\
&+& g \,[G^{\mu\nu} + g\, (F^{\mu\nu} \times \rho) ]\cdot  (\phi_\nu \times C ) 
+ \frac {i}{2} \; g \; \partial^\mu \bar C \cdot (C \times C)
\end{eqnarray}
\begin{eqnarray} 
J^\mu_{ab} &=& - \bar B \cdot (D^\mu \bar C) - \big[F^{\mu\nu} - g\,((G^{\mu\nu} 
+ g\, F^{\mu\nu} \times \rho) \times \rho) 
- m \,\varepsilon^{\mu\nu\eta} \, \phi_\eta \big]  \cdot (D_\nu \bar C)\nonumber\\
&+& g\, [G^{\mu\nu} + g (F^{\mu\nu} \times \rho) ]\cdot  (\phi_\nu \times \bar C ) 
- \frac {i}{2} \;g \; \partial^\mu C \cdot (\bar C \times \bar C)
\end{eqnarray}
These expressions for the Norther's currents can be re-expressed (for the algebraic convenience),
in the following form:  
\begin{eqnarray}
J^\mu_b &\;=\;& B \cdot (D^\mu C) - \partial^\mu B \cdot C 
- \frac {i}{2} \;g \; \partial^\mu \bar C \cdot (C \times C)\nonumber\\
&\;-\;& \partial_\nu [(F^{\mu\nu} - g\, \{(G^{\mu\nu} 
+ g\, F^{\mu\nu} \times \rho) \times \rho\} 
- m \;\varepsilon^{\mu\nu\eta} \; \phi_\eta)\cdot C]
\end{eqnarray}
\begin{eqnarray}
J^\mu_{ab} &\;=\;&  - \bar B \cdot (D^\mu \bar C) + \partial^\mu \bar B \cdot {\bar C} 
+ \frac {i}{2} \;g \; \partial^\mu C \cdot (\bar C \times \bar C)\nonumber\\
&\,\,-\;& \partial_\nu [(F^{\mu\nu} - g\, \{(G^{\mu\nu} + g\, F^{\mu\nu} \times \rho) \times \rho\} 
- m \;\varepsilon^{\mu\nu\eta} \; \phi_\eta)\cdot \bar C]
\end{eqnarray}
Now the proof of  conservation law ($\partial_\mu J^\mu_{(a)b} = 0$) becomes easier and it can be 
confirmed by exploiting the following Euler-Lagrange equations of motion, derived from the 
Lagrangian densities ${\cal L}_B$ and ${\cal L}_{\bar B}$, respectively:
\begin{eqnarray}
&&D_\mu F^{\mu\nu} - g\, D_\mu [(G^{\mu\nu} + g\, F^{\mu\nu}\times \rho) \times \rho] 
+ m \, \varepsilon^{\mu\eta\nu} \, D_\mu \phi_\eta - \partial^\nu B \nonumber\\
&&+ g \,[(G^{\mu\nu} + g\, F^{\mu\nu}\times \rho) \times \phi_\mu]  
- i \, g\, (\partial^\nu \bar C \times C) = 0 \nonumber\\
&&D_\mu [G^{\mu\nu} + g \,(F^{\mu\nu}\times \rho)] 
+ \frac {m}{2} \, \varepsilon^{\mu\eta\nu} \, F_{\mu\eta} = 0 \nonumber\\
&&(G^{\mu\nu} + g\, F^{\mu\nu}\times \rho) \times F_{\mu\nu} = 0 \quad \qquad \partial_\mu (D^\mu C) = 0 
\quad \qquad D_\mu (\partial^\mu \bar C) = 0
\end{eqnarray}
\begin{eqnarray}
&&D_\mu F^{\mu\nu} - g\, D_\mu [(G^{\mu\nu} + g\, F^{\mu\nu}\times \rho) \times \rho] 
+ m \, \varepsilon^{\mu\eta\nu} \, D_\mu \phi_\eta + \partial^\nu \bar B \nonumber\\
&&  + g \,[(G^{\mu\nu} + g\, F^{\mu\nu}\times \rho) \times \phi_\mu]  
+ i \, g\, ( \bar C \times \partial^\nu C) = 0 \nonumber\\
&& D_\mu [G^{\mu\nu} + g \,(F^{\mu\nu}\times \rho) ] 
+ \frac {m}{2} \, \varepsilon^{\mu\eta\nu} \; F_{\mu\eta} = 0 \nonumber\\
&& (G^{\mu\nu} + g\, F^{\mu\nu}\times \rho) \times F_{\mu\nu} = 0 \quad
\qquad \partial_\mu (D^\mu \bar C) = 0 \quad\qquad D_\mu (\partial^\mu C) = 0
\end{eqnarray}
These conserved currents, (35) and (36), lead to the derivation 
of the following expressions for the conserved (anti-)BRST charges $Q_{(a)b} = \int d^2 x \; J^0_{(a)b}$:
\begin{eqnarray}
Q_{ab} = -  \int d^2 x  \Big[\bar B \cdot D^0 \bar C 
- \dot {\bar B} \cdot \bar C 
- \frac{i}{2}\, g\,\dot C \cdot (\bar C \times \bar C) \Big]
\end{eqnarray}
\begin{eqnarray}
Q_b = \int d^2 x  \Big[B \cdot D^0 C - \dot B \cdot C 
- \frac{i}{2}\,g\, \dot {\bar C} \cdot (C \times C) \Big]
\end{eqnarray}
if we exploit the Euler-Lagrange equations of motion, (37) and (38), appropriately. 
It may be mentioned here that these 
charges {\it also} appear in the case of BRST approach to the {\it usual} 4D non-Abelian gauge theory. The form of
expressions (39) and (40) crucially depends on the suitable use of the equations of motion, (37) and (38).
The conserved ($\dot Q_{(a)b} = 0$) and nilpotent (i.e., $Q^2_{(a)b} = 0$) (anti-)BRST charges $Q_{(a)b}$  
are the generators of the transformations (25) and (26), respectively. For instance, it can be checked that 
$s_r \Phi = \pm i\; [\Phi,\; Q_r]_{\pm}$ ($r = b, ab$). Here the field $\Phi$ is the generic field of the theory and 
$(\pm)$ signs, on the square bracket, stand for the (anti)commutator for the generic field $\Phi$ of the
theory being (fermionic)bosonic.

It is interesting to point out that the nilpotent generators $Q_{(a)b}$,
despite producing the nilpotent (anti-)BRST transformations for the basic dynamical fields, are {\it unable} to
generate the (anti-)BRST transformations for the auxiliary field, $\rho$, of the theory. 
Even the requirements of the nilpotency and absolutely anticommuting properties of the proper (anti-)BRST symmetry
transformations do not generate the (anti-)BRST transformations for the auxiliary field,
$\rho$. This is a novel observation in this theory (within the framework of BRST formalism). 
For the usual 4D (non)-Abelian 1-form gauge theory, there are two inputs that lead to the 
derivation of {\it all} the (anti-)BRST  symmetry transformations of {\it all} the relevant fields of the 
specific theory. 
These are (i) the (anti-)BRST charges, $Q_{(a)b}$, as the generators of the nilpotent (anti-)BRST symmetry 
transformations, and
(ii) the requirements of nilpotency and absolute anticommutativity which lead to the derivation of
the (anti-)BRST symmetry transformations for the auxiliary fields of the theory. Thus, the auxiliary field, 
$\rho$, and its transformations are very special.

\section{Ghost charge: BRST algebra}
The Lagrangian densities (30) and (31) remain invariant under the following scale 
transformations for the (anti-)ghost and other basic as well as auxiliary fields of the theory:   
\begin{eqnarray}
&& C \to e^{+ \Sigma}\, C \quad\qquad \bar C \to e^{- \Sigma} \,\bar C  \quad\qquad
 (A_\mu, \phi_\mu, \rho, B, \bar B) \to  e^0 (A_\mu,\, \phi_\mu,\, \rho,\, B,\, \bar B)\qquad
\end{eqnarray}
where $\Sigma$ is a global infinitesimal scale parameter. The ${(+)-}$ signs, in the 
exponents, represent the ghost number of the fields $(C)\bar C$ and the ghost number for the 
rest of the fields (i.e., $A_\mu, \phi_\mu, \rho, B, \bar B$) is equal to zero. 
As a consequence, the latter fields do not transform at all under the ghost-scale transformations. It is 
straightforward to check that the following infinitesimal transformations $(s_g)$, obtained
from (41): 
\begin{eqnarray}
&& s_g C = + \Sigma \, C \quad \qquad s_g \bar C =  - \Sigma \,\bar C \quad \qquad
 s_g(A_\mu,\, \phi_\mu,\, \rho,\, B,\, \bar B) = 0
\end{eqnarray}
are the {\it symmetry} transformations for the Lagrangian densities (30) and (31) because it is 
straightforward to check that $s_g\, {\cal L}_B = 0 = s_g\, {\cal L}_{\bar B}$.

The preceding ghost symmetry transformations, $s_g$, lead to the derivation of Noether's 
conserved current and charge as
\begin{eqnarray}
J^\mu_g &=& i \; \Big[\bar C \cdot D^\mu C - \partial^\mu \bar C \cdot C \Big] \nonumber\\
 Q_g &=& i \int d^2 x \; \Big[\bar C \cdot D^0 C - \dot{\bar C} \cdot C \Big]
\end{eqnarray}
It can be proven that the above ghost charge, $Q_g$, is the generator of (42). 
The nilpotent (anti-)BRST charges $(Q_{(a)b})$ and the ghost charge, $Q_g$,  satisfy the 
following standard BRST algebra:
\begin{eqnarray}
&& Q^2_b = 0 \qquad \qquad Q^2_{ab} = 0 \qquad \qquad i\; [Q_g, \; Q_b] =  Q_b \nonumber\\
&&  i\; [Q_g, \; Q_{ab}] =   - \,Q_{ab} \qquad \qquad Q_g^2 \neq 0\nonumber\\
&&\{Q_b, \;Q_{ab}\} \,=\, Q_b\; Q_{ab} + Q_{ab}\;Q_b = 0
\end{eqnarray}
which shows that the ghost number of the BRST charge is $(+1)$ and that of the anti-BRST charge is $(-1)$.
These statements about the ghost numbers can be checked explicitly by starting with a state 
$ |\psi \rangle_n$ that has the ghost number equal to $n$ (i.e., $ i\, Q_g\, |\psi \rangle_n  
= n \,|\psi \rangle_n $). With this input and the above algebra (44), we can check that the 
ghost numbers of states $ Q_b\, |\psi \rangle_n$ and $ Q_{ab}\, |\psi \rangle_n$ are 
$(n + 1)$ and $(n - 1)$, respectively. Thus, the BRST charge increases the ghost number
by one when it operates on a quantum state. On the other hand, the anti-BRST charge decreases
the above number by one when it acts on the {\it same} quantum state.

\section {Conclusions} 
In our present investigation, we have exploited the {\it usual} classical Yang-Mills gauge symmetry of the JP 
model of 3D massive\footnote{The beauty of the {\it massive} 3D JP model is the observation
that the parity symmetry is not violated  because of the presence of a vector field ($\phi_\mu$) that is endowed
with a parity quantum number opposite to that of the 3D vector gauge (photon) field ($A_\mu$). The existence
of the former field has also been shown from the requirement of the local duality invariance \cite{35}
of Maxwell's equations where this field has been christened as ``axial-vector'' gauge field.}
non-Abelian gauge theory and generalized it onto the (anti-)BRST symmetry 
transformations at the quantum level that are off-shell nilpotent and absolutely 
anticommuting in nature. In this endeavor, the ``augmented" superfield
formalism \cite{25,26,27} (where the HC and GIRs blend together beautifully) has played a decisive role 
in the derivation of the full set of appropriate (anti-)BRST symmetry transformations.

One of the important features of our superfield formulation is the derivation of the (anti-)BRST 
invariant CF condition 
that enables us to obtain the absolutely anticommuting (anti-)BRST symmetry transformations. Thus, 
in addition to the BRST symmetry for the JP model \cite{17},
we have been able to derive the {\it proper} 
anti-BRST symmetry transformations for the sake of completeness. Furthermore, the celebrated CF condition has 
been able to help us in deducing the coupled Lagrangian densities, (30) and (31), that respect the 
these proper (anti-)BRST symmetry transformations for our present theory.

In the context of 4D non-Abelian 1-form gauge theory,  it is well-known that the existence 
of anti-BRST symmetry transformations ($s_{ab}$) is {\it non-trivial}. The off-shell 
nilpotent ($s^2_{a(b)} = 0$) (anti-)BRST  symmetries {\it anticommute} $(s_b \,s_{ab} + s_{ab} \,s_b = 0)$
with each other only on a constrained hypersurface, described by the CF field 
equations, in a 4D Minkowskian  space-time manifold \cite{28}.  It can be also checked, using
the appropriate equations, (25), (26), (39), (40), that for our present theory, the
following equations:
\begin{eqnarray}
s_b Q_{ab} =  - \,i\,  \{Q_b,\; Q_{ab} \} = -\, i\int d^2x \Big[\bar B \cdot \partial^0 \Big(B + \bar B 
\,+ \,i \, g\, (C\times \bar C)\Big)\Big] = 0
\end{eqnarray}
\begin{eqnarray}
s_{ab} Q_b = -\, i\, \{Q_{ab},\; Q_{b} \} = i \int d^2x \Big[B \cdot \partial^0 \Big(B + \bar B 
\,+ \,i \, g\, (C\times \bar C)\Big)\Big] = 0
\end{eqnarray}
are absolutely true on the 3D Minkowski space-time manifold where the CF restriction 
$B + \bar B + i\,g\, (C \times \bar C) = 0$ is satisfied. The absolute anticommutativity 
of the (anti-)BRST symmetries imply the linear independence 
of BRST and anti-BRST symmetry transformations. The mathematical basis for the independence of 
the BRST and anti-BRST symmetries is  encoded in the concept of  gerbes that has been discussed
in our earlier works \cite{18,19} and illustrated geometrically in our Appendix B where we pin-point
the existence of CF condition.

To obtain the full set of proper (anti-)BRST symmetry transformations, we are theoretically 
compelled to go beyond the HC and exploit the suitable GIRs to deduce the proper (anti-)BRST
symmetries. This is a novel feature of this model. Furthermore, as it turns out, the auxiliary 
field, $\rho$, is {\it not} like the other auxiliary  (e.g., Nakanishi-Lautrup) fields of the theory because 
its (anti-)BRST symmetry transformations do {\it not} arise from the requirements of nilpotency 
and absolute anticommutativity of the (anti-)BRST symmetry transformations. This observation is 
{\it also} a novel feature of our present theory. The good thing about our present augmented 
superfield formalism \cite{25,26,27} is that 
it leads to the precise
derivation of the proper (anti-)BRST symmetry transformations associated with this {\it special} 
auxiliary field, $\rho$, as well, for our present 3D non-Abelian theory.

It would be very nice endeavor to exploit the NYM gauge transformations, (3), within the framework 
of our superfield formalism and obtain the novel results connected with the nilpotent (anti-)BRST symmetry 
transformations that emerge from it. In fact, two of us (SG and RK), have already  
obtained the proper (anti-)BRST symmetry transformations corresponding to NYM symmetries by 
taking into account the following restrictions on the (super)fields  \cite{33}:
\begin{eqnarray}
\tilde d \tilde \phi^{(1)} + i\, g\, (\tilde A^{(1)} \wedge \tilde \phi^{(1)}) 
+ i\, g\, (\tilde \phi^{(1)}
\wedge \tilde A^{(1)}) - i\, g\, (\tilde F^{(2)} \wedge \tilde \rho) 
+ i\, g\, ( \tilde \rho \wedge \tilde F^{(2)}) \;\;\quad\quad\nonumber\\
 =\; d \phi^{(1)} + i\, g\, (A^{(1)} \wedge \phi^{(1)}) + i\, g\, (\phi^{(1)} 
\wedge  A^{(1)}) - i\, g\, (F^{(2)} \wedge \rho)
 +\, i\, g\, (\rho \wedge F^{(2)})
\end{eqnarray}
where the symbols carry their standard meanings (as discussed in our present text). 
It would also be exciting to take the combination 
of the local  Yang-Mills and NYM gauge transformations {\it together} and obtain the full set of proper (anti-)BRST
symmetry transformations, relevant coupled Lagrangian densities and exact 
(conserved,  nilpotent and absolutely anticommuting) (anti-)BRST charges that generate the
proper and full set of (anti-)BRST symmetry transformations for the JP model. Our present work and the results of
ref. 33 would be the limiting cases of this general approach to the JP model 
(where the Yang-Mills and NYM symmetries would blend together). In this context, it is to be noted that
in a very recent interesting publication \cite{36}, it has been mentioned that
the 3D JP model would turn out to be ultraviolet finite and renormalizable. In 
this work \cite{36}, the proper gauge-fixed BRST invariant Lagrangian density
has been found. We plan to tap some of the inputs from \cite{36} to apply our superfield 
formalism so hat we could obtain the proper (anti-)BRST symmetries corresponding to the
Yang-Mills and NYM symmetries {\it together}. We hope to report about 
it in our future publication \cite{37}.

\section*{Acknowledgments}
SG and RK would like to gratefully acknowledge the financial
support from CSIR and UGC, New Delhi, Government of India, respectively.

\section*{Appendix A: \\
Nilpotency and (anti-)BRST invariance: superfield technique}

The nilpotency and anticommutativity of the (anti-)BRST symmetry transformations (and
corresponding generators) can be proven, in a simple and elegant manner, by exploiting
the potential and power of superfield formalism. As has been pointed out earlier, it can 
be checked, from the expansions in (9) and (15), that 
%\begin{eqnarray}
\[s_b \Psi (x)  =  \frac {\partial}{\partial \bar \theta} \, 
\tilde \Psi^{(h, g)}(x,\theta,\bar\theta)\Big|_{\theta = 0}\qquad \qquad
s_{ab} \Psi (x)  =   \frac {\partial}{\partial \theta} 
\, \tilde \Psi^{(h, g)}(x,\theta,\bar\theta)\Big|_{\bar \theta = 0}\]
\[s_b\, s_{ab} \Psi (x)  =  \frac {\partial}{\partial \bar \theta} \, \frac {\partial}{\partial \theta} 
\, \tilde \Psi^{(h, g)} (x,\theta,\bar\theta) \eqno{(A.1)}\]
%\end{eqnarray}
where $\Psi (x)$ is the generic 3D field of the theory and $\tilde \Psi^{(h, g)} (x,\theta,\bar\theta)$ 
is the corresponding
superfields (defined on the $(3, 2)$-dimensional supermanifold and expanded after the 
application of HC and (or) GIRs).
As a consequence, the off-shell nilpotency of the (anti-)BRST transformations is captured in the 
nilpotency ($\partial_\theta^2 = 0, \; \partial_{\bar \theta}^2 = 0$) of the Grassmannian directions
($\partial_\theta, \partial_{\bar \theta} $). In an exactly similar fashion, it can be checked that
%\begin{eqnarray}
\[ \Big(\frac {\partial}{\partial \theta} \; \frac {\partial}{\partial \bar \theta} +
\frac {\partial}{\partial \bar \theta} \; \frac {\partial}{\partial \theta} \Big) \; \tilde \Psi^{(h, g)} 
(x, \theta, \bar \theta)  =  0\eqno{(A.2)}\]
%\end{eqnarray}
encodes the absolute anticommutativity $(s_b\, s_{ab} + s_{ab}\, s_b = 0)$ of the (anti-)BRST
symmetry transformations in 3D space-time for the JP model of massive theory.

That the (anti-)BRST charges are {\it also} nilpotent of order two, can be captured in the following 
expressions within the framework of superfield formalism:
%\begin{eqnarray}
\[ Q_b = \frac {\partial}{\partial \bar \theta} \; \int d^2 x \;
\Big[B(x) \cdot \tilde B_0^{(h)} (x, \theta, \bar\theta) + i \; {\dot{\tilde {\bar F}}}^{(h)}(x, \theta, \bar\theta)
\cdot \tilde F^{(h)} (x, \theta, \bar\theta) \Big] \bigg|_{\theta = 0} \qquad\;\]
\[\equiv \int d^2 x  \int d \bar\theta \; 
\Big[B(x) \cdot \tilde B_0^{(h)} (x, \theta, \bar\theta) + i \; {\dot{\tilde {\bar F}}}^{(h)}(x, \theta, \bar\theta)
\cdot \tilde F^{(h)} (x, \theta, \bar\theta ) \Big]\bigg|_{\theta = 0} \eqno{(A.3)}\]
%\end{eqnarray}
Written in the ordinary 3D space-time, the above expansions imply
%\begin{eqnarray}
\[Q_b = \int d^2 x\; s_b \; \Big[B(x) \cdot A_0 (x) + i \; \dot{\bar C} (x) \cdot C(x) \Big] \eqno{(A.4)} \]
%\end{eqnarray}
As a consequence, it is clear that $s_b\, Q_b = - \,i\, \{Q_b,\; Q_b \} = 0$ because of the nilpotency 
($s_b^2 = 0$) of the BRST transformations ($s_b$). In the language of superfield and superspace 
variables (with the inputs from the nilpotency $(\partial^2_{\bar \theta} = 0)$ of derivative 
$\partial_{\bar \theta}$)
%\begin{eqnarray}
\[ \frac {\partial}{\partial \bar\theta} \; Q_b \bigg|_{\theta = 0}  =  0 
\quad \Longrightarrow \quad Q_b^2  =  0 \eqno{(A.5)}\]
%\end{eqnarray} 
There is an alternative way to express BRST charge (that is valid only on the constrained surface
in 3D space-time manifold where CF-condition $[B\, +\, \bar B \,+ \,i\, g \,(C \times \bar C) = 0]$ is satisfied). 
This is given, within the framework of superfield formalism, as
%\begin{eqnarray}
\[Q_b  = i\, \frac {\partial} {\partial \bar \theta} \; \frac {\partial} {\partial  \theta} \int
d^2 x \; \Big[ \tilde B_0^{(h)} (x, \theta, \bar \theta) \cdot \tilde F^{(h)} (x, \theta, \bar 
\theta) \Big] \]
\[ \equiv \int d^2 x \; s_b\, s_{ab} \; \Big(i\; A_0 (x) \cdot C(x)\Big) \qquad\qquad\eqno{(A.6)} \]
%\end{eqnarray}
The off-shell nilpotency of the BRST charge is, once again, proven by the nilpotency of Grassmannian derivative
$\partial_{\bar \theta}$ (i.e., $ \partial_{\bar \theta}^2 = 0$) and the off-shell nilpotency 
($s_b^2 = 0$) of the BRST transformations $s_b$. Thus, we note that the nilpotency of BRST 
charge is encoded in $\partial_{\bar \theta}^2 = 0$ and $s_b^2 = 0$ when it is expressed in 
terms of the Grassmann derivative, $\partial_{\bar \theta}$, on the $(3, 2)$-dimensional supermanifold and in terms of 
the proper BRST transformations, $s_b$, existing in the 3D ordinary space-time.

As we have expressed the BRST charge, $Q_b$, in terms of the superfields (obtained after the 
application of HC and (or) GIRs), we can also express the anti-BRST charge, $Q_{ab}$, in the following two different 
ways:
%\begin{eqnarray}
\[Q_{ab}  =  \frac {\partial}{\partial \theta} \; \int d^2 x \;
\Big[\bar B(x) \cdot \tilde B_0^{(h)} (x, \theta, \bar\theta) + i \; \dot{\tilde{F}}^{(h)}(x, \theta, \bar\theta)
\cdot \tilde {\bar F}^{(h)} (x, \theta, \bar\theta) \Big]\bigg|_{\bar \theta = 0} \]
\[Q_{ab}  = i \; \frac {\partial}{\partial \theta} \frac {\partial}{\partial \bar \theta} \; 
\int d^2 x \;  \Big[ \tilde B_0^{(h)} (x, \theta, \bar\theta) \cdot 
\tilde {\bar F}^{(h)} (x, \theta, \bar\theta) \Big] \qquad\quad\qquad\;\;\qquad\quad\eqno{(A.7)}\]
%\end{eqnarray}
where the second expression is true only on the 3D constrained hypersurface where CF condition 
$B + \bar B + i\, g\, (C \times \bar C) = 0$ is satisfied. The nilpotency ($\partial_\theta^2 = 0$) 
of the Grassmannian derivative $\partial_\theta$ ensures that $Q_{ab}^2 = 0$ because it can be seen
that $\partial_\theta Q_{ab} = 0 $. The above expression can be expressed, in the 3D ordinary space, as
%\begin{eqnarray}
\[Q_{ab} = \int d^2 x \; s_{ab}\;  \Big [ \bar B(x) \cdot A_0(x) + i \; \dot C (x) \cdot \bar C (x) \Big] \]
\[\equiv i\; \int s_{ab} s_b \; \Big( A_0 (x) \cdot \bar C(x) \Big) \qquad\qquad\;\;\, \eqno{(A.8)}\]
%\end{eqnarray}
The nilpotency of $Q_{ab}$ is captured in the equation $s_{ab} Q_{ab} = - i \{Q_{ab}, Q_{ab} \} = 0$
because we know, from the anti-BRST symmetry transformations (25), that $s_{ab}^2 = 0$.

The (anti-)BRST invariance and the equivalence of the coupled Lagrangian densities can {\it also}
be captured in the language of superfield formalism. With this end in mind, it can be checked 
that the starting Lagrangian density, ${\cal L}_0$, can be generalized onto the  $(3, 2)$-dimensional 
supermanifold as 
%\begin{eqnarray}
\[{\cal L}_0 \to \tilde {\cal L}_0 = - \frac{1}{4}\, \tilde F^{\mu\nu (h)} 
\cdot \tilde F_{\mu\nu}^{(h)} - \frac{1}{4}\, \Big[\tilde G^{\mu\nu (h, g)} 
+ g \, \tilde F^{\mu\nu (h)} \times \tilde \rho^{(h,g)} \Big]
\cdot \Big[\tilde G_{\mu\nu}^{(h, g)} + g \, \tilde F_{\mu\nu}^{(h)} \times \tilde \rho^{(h,g)} \Big] \]
\[+ \,\frac {m}{2}\;\varepsilon^{\mu\nu\eta} \; \tilde F_{\mu\nu}^{(h)} 
\cdot \tilde \phi_\eta^{(h,g)} \qquad\qquad\qquad\qquad\qquad\qquad\qquad\qquad\qquad\;\; \eqno{(A.9)}\]
%\end{eqnarray}
Theis expression for $\tilde{\cal L}_0$ is, however, independent of the Grassmannian variables because 
of our clear discussion in Sect. 3 and Sect. 4. As a consequence, we have
%\begin{eqnarray}
\[\frac {\partial}{\partial \bar \theta} \; \tilde {\cal L}_0\bigg|_{\theta = 0} \; =  \;
\frac {\partial}{\partial  \theta} \; \tilde {\cal L}_0\bigg|_{\bar \theta = 0} \; =\;  0 
\quad \Longleftrightarrow \quad s_{(a)b} \; {\cal L}_0 = 0 \eqno{(A.10)}\]
%\end{eqnarray}
Thus, we conclude that
this equation captures the (anti-)BRST invariance of the starting Lagrangian density, ${\cal L}_0$,
in terms of the superfield and superspace variables.

The (anti-)BRST invariant coupled Lagrangian densities can be generalized onto the $(3, 2)$-dimensional
supermanifold as
%\begin{eqnarray}
\[{\cal L}_{\bar B} \to \tilde {\cal L}_{\bar B} =  \tilde {\cal L}_0 - \frac {\partial} 
{\partial \theta}\, \frac {\partial} {\partial \bar \theta}\,  \Big [ \frac {i} {2} \, \tilde B_\mu^{(h)} 
\cdot \tilde B^{\mu (h)} + \tilde F^{(h)} \cdot \tilde {\bar F}^{(h)} 
+ \frac {1} {2} \, \tilde \phi_\mu^{(h,g)} \cdot \tilde \phi^{\mu (h,g)} \Big] \qquad\]
\[{\cal L}_{B} \to \tilde {\cal L}_{B}  = \tilde {\cal L}_0 + \frac {\partial} 
{\partial \bar \theta}\, \frac {\partial} {\partial  \theta}\,  \Big [ \frac {i} {2} \, \tilde B_\mu^{(h)} 
\cdot \tilde B^{\mu (h)} + \tilde F^{(h)} \cdot \tilde {\bar F}^{(h)} 
+ \frac {1} {2} \, \tilde \phi_\mu^{(h,g)} \cdot \tilde \phi^{\mu (h,g)} \Big] \eqno{(A.11)}\]
%\end{eqnarray}
The nilpotency ($\partial^2_\theta = \partial^2_{\bar \theta} = 0$) of the Grassmannian derivatives 
($\partial_\theta,  \partial_{\bar \theta} $) and Grassmannian independence of $\tilde {\cal L}_0$,
lead to the following:
%\begin{eqnarray}
\[\frac {\partial}{\partial \bar \theta}\;  \tilde{\cal L}_B \bigg|_{\theta = 0}\; = \; 0 
\; \Longleftrightarrow \; s_b \; {\cal L}_B = 0 \]
\[ \frac {\partial}{\partial \theta}\;  \tilde{\cal L}_{\bar B}\bigg|_{\bar \theta = 0} \; = \; 0 
\; \Longleftrightarrow \; s_{ab} \; {\cal L}_{\bar B} = 0 \eqno{(A.12)}\]
%\end{eqnarray}
The equivalence of the coupled Lagrangian densities, ${\cal L}_B$ and ${\cal L}_{\bar B}$, is 
captured in the proof that $(\partial/ \partial \theta)
 \; \tilde {\cal L}_B \big|_{\bar \theta = 0} = 0, \;  (\partial/ \partial \bar \theta)
 \; \tilde {\cal L}_{\bar B} \big|_{\theta = 0} = 0$ because of the nilpotency ($\partial_\theta^2 = 0, \partial_{\bar\theta}^2 = 0$)
and anticommutativity of the Grassmannian
 derivatives (i.e. $\partial_\theta \,\partial_{\bar\theta}
+ \partial_{\bar\theta}\, \partial_{\theta} = 0$).

\section*{Appendix B:\\
 (Anti-)BRST symmetries, CF condition and gerbes:
physical interpretation and geometrical meaning}

In this Appendix, we provide a clear geometrical interpretation for the existence of CF condition in the 
language of (anti-)BRST symmetry transformations on the non-Abelian 1-form gauge field 
$A_\mu \equiv A_\mu \cdot T$ and associated
(anti-)ghost fields. It is evident, from a close look at the off-shell nilpotent and
absolutely anticommuting (anti-)BRST transformations, (25), (26) and (27),
that the  schematic diagram Fig. A1   
\begin{figure}[b]
\centerline{\includegraphics[width=12cm]{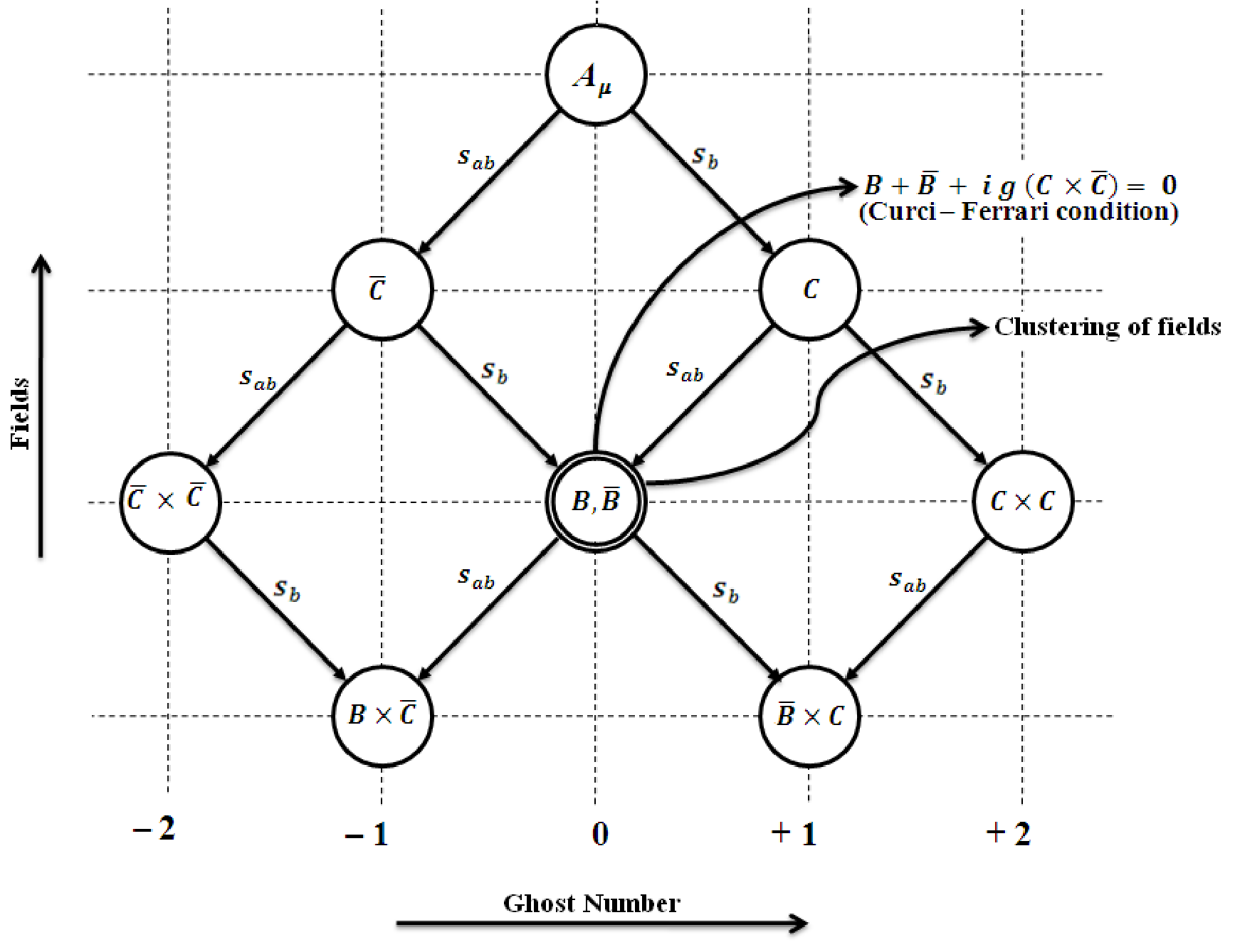}}
\caption{(Anti-)BRST symmetry transformations for the  non-Abelian 1-form  gauge field, 
associated (anti-)ghost fields and emergence of CF condition.}
\end{figure}
captures these transformations geometrically. We note that, in the whole diagram, there is a {\it single point},
at the ghost number {\it zero}, where there is a clustering of  the auxiliary fields $B \equiv B \cdot T$ and
$\bar B \equiv \bar B \cdot T$, which emerge from the (anti-)BRST symmetry transformations on the
ghost fields $C \equiv C \cdot T$ and anti-ghost fields $\bar C \equiv \bar C \cdot T$, respectively.
This is the place where the CF condition exists as it connects these auxiliary Nakanishi-Lautrup
fields, $B$ and $\bar B$, with the ghost number {\it zero} object constructed from the fermionic (anti-)ghost fields
$(\bar C)C$ of the theory  (i.e., $B + \bar B + i\;g\;(C \times \bar C) = 0$).

In our earlier works on the 4D Abelian 2-form and 6D 3-form gauge theories, \cite{18,19}  we have explained
the existence and emergence of the (anti-)BRST invariant CF-type restrictions  by exploiting the
fundamental notions of  geometry and group theory from pure mathematics. As a warm-up exercise in ref. 19,
we have also considered the existence of CF condition for the 4D non-Abelian 1-form gauge theory within
the framework of BRST formalism by taking the help of concepts from  pure mathematics.
In our present endeavor,
we claim that, for any arbitrary $p$-form gauge theory within the framework of BRST formalism,
it is the symmetry transformations that would provide us the clue for the existence of CF-type
conditions in the theory. In fact, wherever, in the above type of diagram (Fig. A1), there is
clustering  of different variety of fields at a particular ghost number, there will  emerge a CF-type
condition (which will be mathematically connected with the idea of gerbes \cite{18,19}). 
Thus, physically, we interpret the
mathematical object gerbes as some artifact that connects the fields, which cluster at a particular
ghost number (in the diagram like Fig. A1 for a given theory) and they basically prove the linear
independence (i.e. the absolute anticommutativity) of the BRST and anti-BRST symmetries.

The hallmark of a gauge theory, at the classical level, is the existence of first-class 
constraints on them in the language of Dirac's prescription for classification scheme. 
On the other hand, one of the decisive features of a gauge theory, at the quantum level, 
is the existence of CF-type conditions (which are mathematically  backed by the idea of 
gerbes) within the realm of BRST approach.

\end{document}